\documentclass[preprint,12pt]{elsarticle}
\usepackage{amsmath}
\usepackage {graphics}
\usepackage{graphicx}
\usepackage{color}
\usepackage[nodots,nocompress]{numcompress}
\usepackage{amssymb}
\usepackage{nomencl}

\begin{document}
\begin{frontmatter}

\title{EXPLICIT MAPPING OF ACOUSTIC REGIMES FOR WIND INSTRUMENTS}

\author{Samy Missoum\fnref{fn1} }
    \address{
	Department of Aerospace and Mechanical Engineering\\
	University of Arizona. Tucson, Arizona 85721\\
    Email: smissoum@email.arizona.edu
    }

\author{Christophe Vergez, Jean-Baptiste Doc} 
    \address{Laboratoire de M\'ecanique et d'Acoustique. LMA, CNRS, UPR 7051.\\
         Centrale Marseille \\
	F-13402 Marseille Cedex 20. France\\
	Email: \{vergez, jbdoc\}@lma.cnrs-mrs.fr
    }

\begin{abstract}
\textcolor{black}{\it This paper proposes a methodology to map the various acoustic regimes of wind instruments. 
The maps can be generated in a multi-dimensional space consisting of design, control parameters, and initial conditions. The boundaries of the maps are obtained explicitly in terms of the parameters using a support vector machine (SVM) classifier as well as a dedicated adaptive sampling scheme. The approach is demonstrated on a simplified clarinet model for which several maps are generated based on different criteria. Examples of computation of the probability of occurrence of a specific acoustic regime are also provided. In addition, the approach is demonstrated on a design optimization example for optimal intonation.}
\end{abstract}

\fntext[fn1]{Corresponding Author}
\end{frontmatter}


%

						\section{INTRODUCTION}

Wind instruments (e.g., clarinets, oboes, trumpets) are nonlinear systems characterized by the presence of various acoustic regimes~\cite{Mcintyre1983,ClarinetLogic}. The control of  these regimes is typically tedious as exemplified by the difficulty for beginners to successfully produce a sound. In this case, the musician tries to reach the conditions discriminating the two following regimes: the occurrence or the lack of self-sustained oscillations in the instrument's resonator (i.e., a sound is produced or not).
Beyond this simplified description, playing an instrument requires the selection of specific regimes (e.g., a given note) among many possible ones~\cite{wilson74}.

Despite recent improvements in the understanding of the physics and the modeling of wind instruments~\cite{chaigne2008acoustique}, there are still many open questions related to how the various regimes are reached. More specifically, it is not yet understood how the coupling between design and control parameters of the instrument will trigger a given regime. Such an understanding could help to not only understand the important features of a ``good instrument" but also help, in the long term, modify the design of instruments for improved playability.

In order to understand the coupling between the various parameters and the acoustic behavior of the instrument, this paper proposes to generate maps defining the regions where specific acoustic regimes are reached. These maps are generated in a space of parameters that are known to govern the acoustic behavior of the instruments (e.g., mouth pressure and reed channel opening). The ability to construct such maps to associate specific acoustic behaviors with regions of the parameter space is of crucial importance for the analysis and design of wind instruments.
In general, such a mapping is non intuitive and cannot be obtained through analytical developments arbitrarily far from the oscillation threshold.
Alternative approach consists in performing the numerical continuation of solution branches~\cite{karkar2013b,cochelin09,doedel09} but only static or periodic solution branches are tractable~\cite{karkar2012,terrien2013a}. More importantly, there are several parameters influencing the acoustic behavior of the instrument thus requiring the construction of a map in a multidimensional space. 
For this purpose, the most intuitive approach is the construction of maps using a cartesian grid ~\cite{inacio2008,inacio2007,guettler2002}.
However,  precisely identifying the boundaries between various regions of the map requires a large amount of simulation calls which increases exponentially as the dimensionality of the parameter space increases.
For this reason, a dedicated approach such as the one proposed in this article is needed.

In this work, the map boundaries are constructed using a Support Vector Machine (SVM) \cite{cristianini2006introduction,prop53} classifier. SVM is a so-called ``machine learning" technique and defines an ``optimal"  boundary between two sets of data points differentiated by their ``classes" (e.g., pass or fail ). That is, an SVM requires an initial training step such as a design of experiments \cite{santner2003design,prop27}.  The advantage of SVM stems from the construction of a boundary defined explicitly in terms of the parameters. In addition, the boundary generated can be highly nonlinear and defined in a high dimensional space. Because, the initial SVM constructed from a design of experiment might not be accurate, it is subsequently refined using a dedicated adaptive sampling scheme developed by the first author \cite{basudharimproved}. 

The flexibility of the approach allows one to study the regimes in a hybrid space including design, control parameters as well as initial conditions. The technique also enables the efficient propagation of uncertainties to estimate the probability of belonging to a specific regime \cite{Basudhar2009}.  In addition, because SVM is a classification technique, it is not sensitive to discontinuous behaviors. This is an essential aspect since this type of behavior can occur in wind instruments (e.g., discontinuities in playing frequencies due to changes in register). Another advantage of SVM is the possibility of handling several classification criteria simultaneously (e.g., frequency and amplitude).

\textcolor{black}{The aim of this paper is to demonstrate the flexibility and advantages of the proposed methodology rather than to contribute to the understanding of the physics of a particular instrument.
The capabilities of the new mapping approach are demonstrated through various numerical experiments based on the simplified reed instrument model.  Two- and three-dimensional maps are presented for various choices of parameters, including initial conditions.  In addition, examples of calculation of probability of obtaining a given frequency are provided. Finally, the approach is applied to a simple design optimization example for intonation.}

\textcolor{black}{The paper is structured as follows. 
The first section provides a background on the construction of the maps using SVM.
The second section introduces the basic physics of a clarinet as well as the corresponding simplified model.
Finally, the last section provides a set of numerical results in order to highlight the capabilities and potential of the methodology when applied to musical acoustics.}

	 \section{EXPLICIT DESIGN SPACE DECOMPOSITION \label{sec:EDSD}}

\subsection{Support Vector Machines (SVMs)}

In order to build maps, a technique referred to as explicit design space decomposition (EDSD) \cite{basudharimproved,Basudhar2009}, is used. The basic idea is to construct the boundaries of an $n$-dimensional map using a Support Vector Machine (SVM) \cite{prop53,gunn}, which provides an explicit expression of the boundary in terms of the chosen parameters. 

\begin{figure}[h]
\centering
\includegraphics[width=0.45\textwidth]{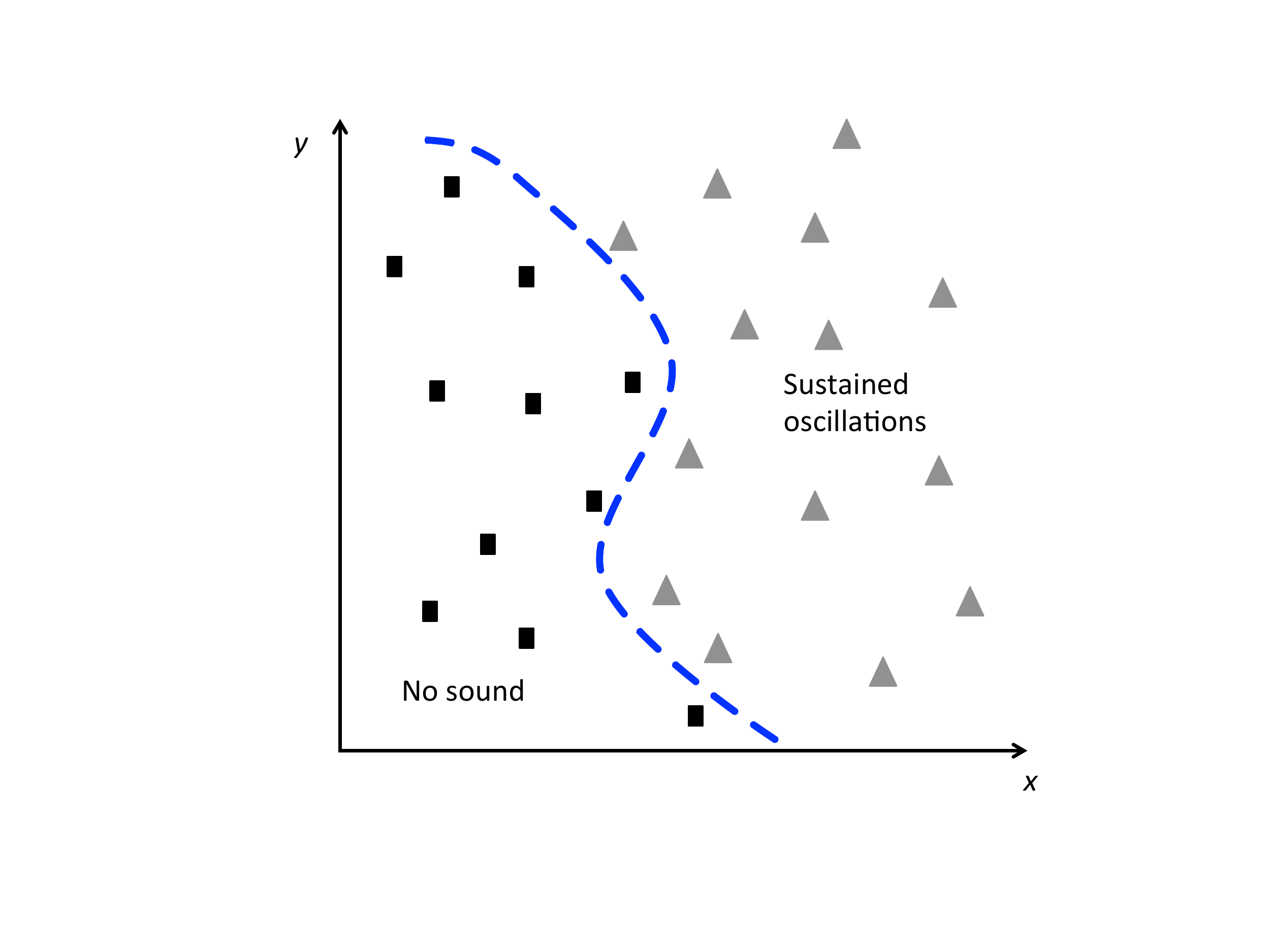}
\caption{Schematic depiction of an SVM boundary classifying sound producing configurations and soundless ones.\label{fig:conceptSVM}
}
\end{figure}

SVM is a machine learning technique that is widely used for classification. In optimization and reliability assessment, SVMs \textcolor{black}{are} used to approximate highly nonlinear constraints and limit-state functions. The most important features of SVMs are their ability to handle multiple criteria using a single classifier, to be insensitive to discontinuities \cite{missoum2010reliability}, and to be computationally very efficient. The ability of an SVM to handle discontinuities is essential in the case of sudden changes of acoustic regimes (e.g., change of register). A schematic example of an SVM separating configurations with sounds and without sound is depicted in Figure \ref{fig:conceptSVM}. 

An SVM defines an explicit boundary that separates samples belonging to two classes labeled as $+1$ and $-1$. Given a set of $N$ training samples $\mathbf{x}_{i}$ in an $n$-dimensional space and the corresponding class labels \textcolor{black}{$y_i=\pm 1$}, an SVM boundary is given as:
\begin{equation}
\label{decisionfinal}
s(\mathbf{x})=b+\sum_{i=1}^{N}\lambda_{i}y_{i}K(\bf{x}_{i},\bf{x})=0
\end{equation}
where $b$ is a scalar referred to as the bias, $\lambda_{i}$ are Lagrange multipliers , and $K$ is a kernel function. \textcolor{black}{The Lagrange multipliers and the bias are solved for using a quadratic programming optimization problem. The optimization problem is formulated so as to maximize the margin between samples of both classes \cite{prop53,gunn}}. The training samples for which the Lagrange multipliers are non-zero are referred to as the \textit{support vectors}.  The predicted class of any arbitrary point $\mathbf{x}$ is given by the sign of $s(\mathbf{x})$. 

The kernel function $K$ in Equation~\eqref{decisionfinal} can have several forms, such as polynomial or Gaussian radial basis kernel. The Gaussian kernel (Equation~\eqref{eqgaussian}) is used in this article:
\begin{equation}\label{eqgaussian}
K(\mathbf{x}_{i},\mathbf{x}_{j})=\exp{ \left( - \frac{\left|\left|\mathbf{x}_i-\mathbf{x}_j \right|\right|^2 }{2\sigma^2}            \right)},
\end{equation}
where $\sigma$ is the width parameter.

An initial approximation of the map is obtained using a design of experiments (DOE) \cite{prop27,prop66,SacksNov.1989} such as Latin Hypercube Sampling (LHS) or Central Voronoi Tessellation (CVT) \cite{lcvt}. These DOE techniques are tailored so as to provide information over the whole space using a reasonable number of samples in higher dimensions.  From an implementation point of view, it is important that the variables be scaled for the construction of the SVM. 

\subsection{Refinement of the SVM boundary. Adaptive sampling.}

\begin{figure}[h]
\textcolor{black}{
\centering
\includegraphics[width=0.7\textwidth]{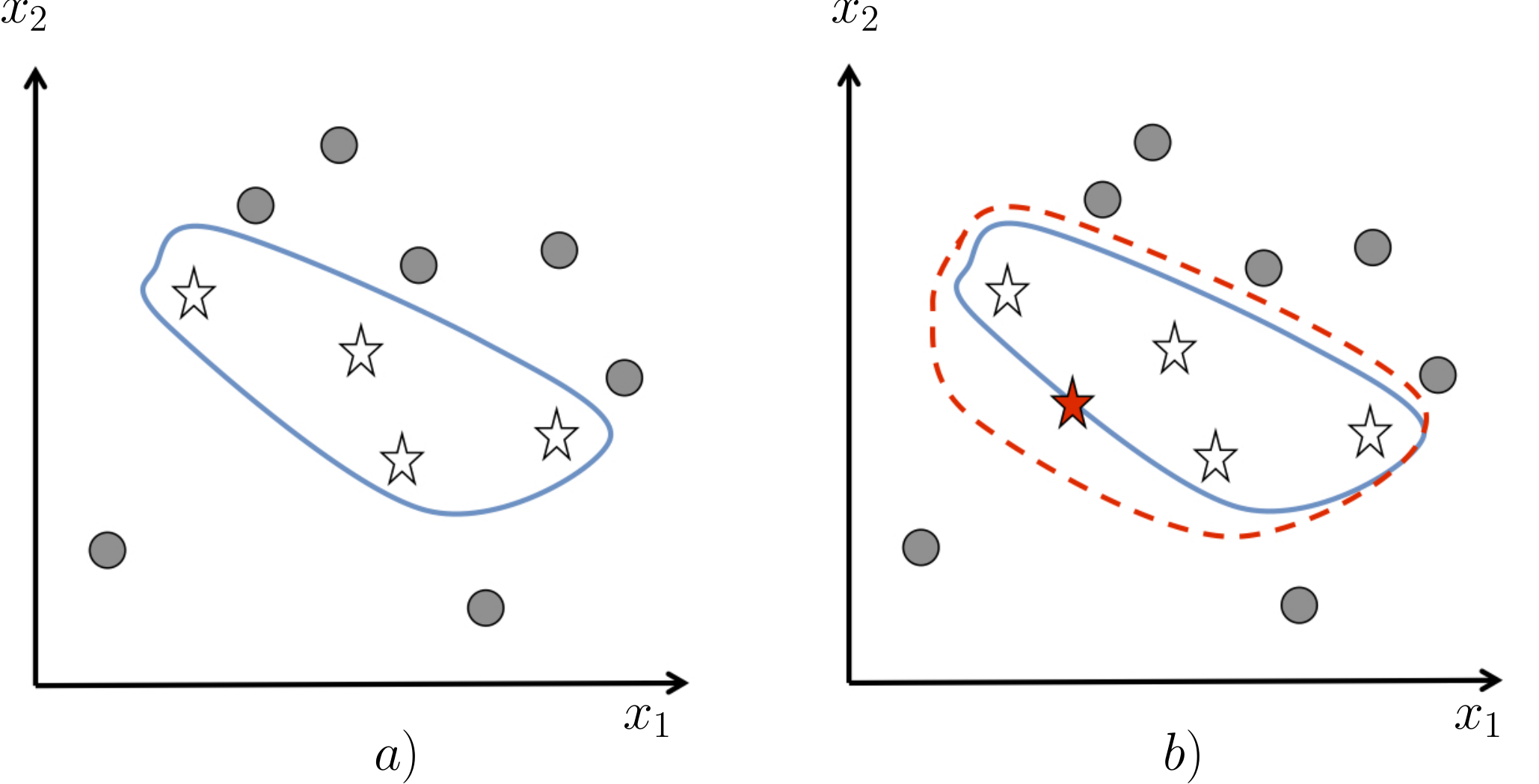}
\caption{Schematic representation of the EDSD adaptive sampling scheme before (a) and after adding one sample (b). The addition of a sample (red star) in the sparse region of the space on the SVM is the basis of EDSD. The red dashed boundary corresponds to the updated SVM. \label{fig:adapt}}}
\end{figure}

The initial approximation of the boundary using a DOE might not be accurate and needs to be refined while maintaining a reasonable number of calls to the acoustic model. This refinement is performed using an adaptive sampling scheme that was developed by the first author and his team and is described in \cite{basudharimproved}. The basis of the scheme is to select a sample in the sparse regions of the space (i.e., as far away as possible from existing sample) and also in the regions of highest probability of misclassification by the SVM. The latter criterion is obtained by locating the samples on the SVM.  In order to illustrate the foundation of the algorithm, Figure \ref{fig:adapt} depicts an SVM constructed from two classes of samples (circles and stars) which is subsequently refined using an adaptive sample located in the most sparse region on the boundary.

Formally, the sample is found by solving the following optimization problem:
\begin{align}
\label{GAMAXMIN}
    \max_{\mathbf{x}} &\quad \left|\left|{\mathbf{x}}-{\mathbf{x}}_{nearest}\right|\right|\\
    s.t.&\quad s(\mathbf{x}) = 0 \notag\\
&\quad   \mathbf{x}_{min} \leq \mathbf{x} \leq \mathbf{x}_{max} 
\end{align}
where ${\mathbf{x}}_{nearest}$ is the closest training sample (i.e., the one with minimum distance). This is a global (non-smooth) ``max-min" optimization problem that can either be solved using a global optimizer such as a Genetic Algorithm or using a local optimizer with different starting points. A gradient-based technique such as Sequential Quadratic Programming (SQP) can be used by reformulating the problem into a differentiable one as detailed in  \cite{basudharimproved}.

\subsection{Probability estimates}
\label{sec:prob}
Once an SVM boundary is constructed, the probability \textcolor{black}{that a configuration belongs to a specific class} (e.g., belongs to the region of the space corresponding to self-sustained oscillations) can be obtained through Monte-Carlo simulations \cite{prop56}. \textcolor{black}{Given the probability density functions  of the parameters (e.g., the blowing pressure follows a normal distribution),} the probability $P_f$ of belonging to the ``positive" class is approximated using $N_{MC}$ samples $\mathbf{p}_i$:

\begin{equation}
\label{eq:MCprob}
P_f=\frac{1}{N_{MC}}\sum_{i=1}^{N_{MC}}  I_s(\mathbf{p}_i), 
\end{equation}

where $I_s$ is the indicator function defined as:
\[ I_s= \left\{
  \begin{array}{l l}
    0 & \quad \text{if $s(\mathbf{p}_i) \leq 0$}\\
    1 & \quad \text{if $s(\mathbf{p}_i) > 0$}
  \end{array} \right.\]

			\section{\textcolor{black}{CASE STUDY: THE CLARINET} \label{model}}

\textcolor{black}{In order to demonstrate the proposed methodology for the mapping of acoustic regimes and provide the reader with the necessary background, this section focuses on clarinets.
It is an arbitrary choice and any other musical instrument  could have been chosen.
An explanation of the physics involved in the production of sound is first provided.
A simplified clarinet model is subsequently described. }

\subsection{The physics - Fundamental equations}

\begin{figure}[h]
\centering
\includegraphics[width=\columnwidth]{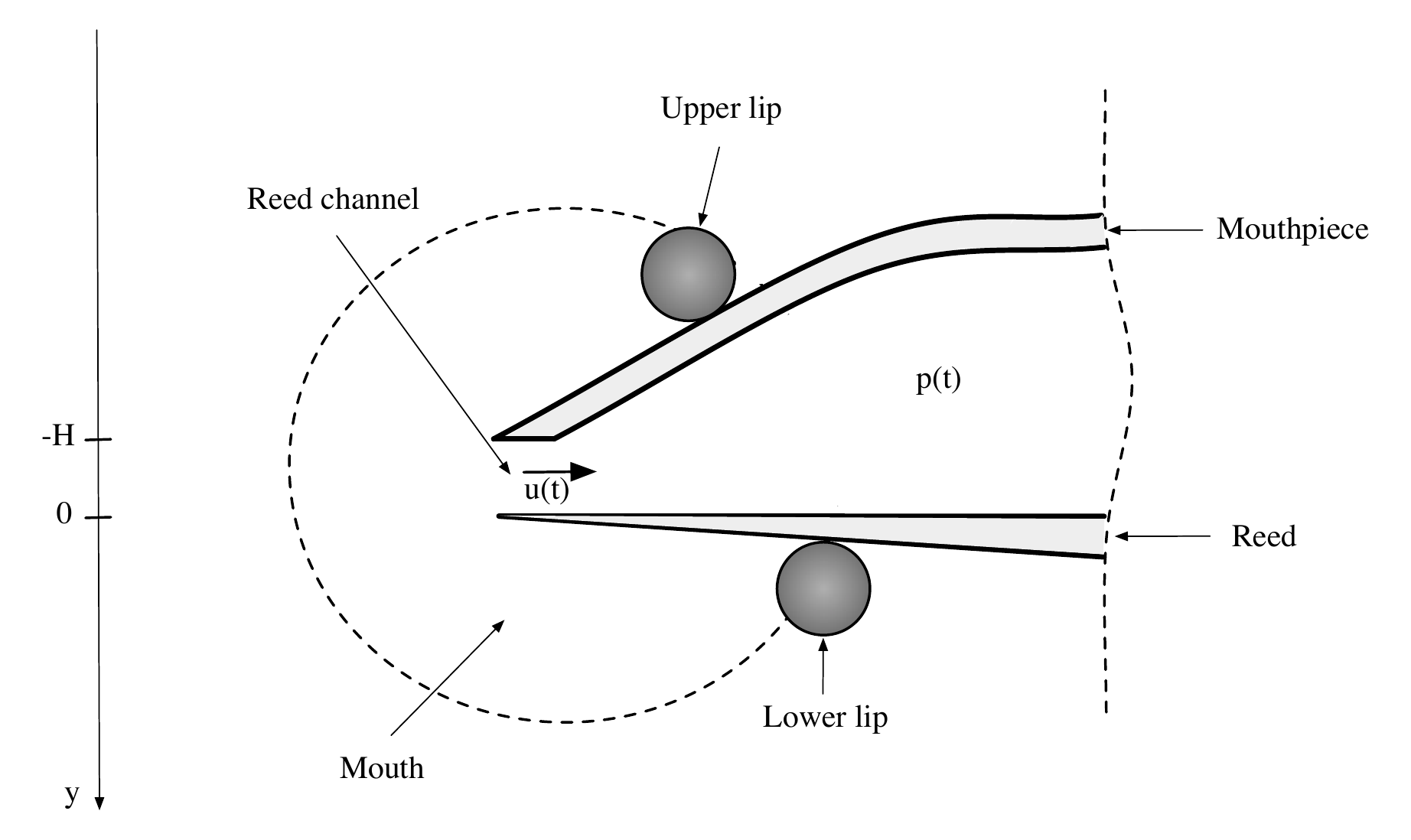}
	\caption{Schematic representation of a clarinet.}
\label{fig_clari_scheme}
\end{figure}

By blowing air into the instrument through the reed channel (Figure \ref{fig_clari_scheme}), the player destabilizes the reed from its rest position. The oscillations of the reed modulate the incoming air flow $\tilde{u}(t)$, thus exciting the acoustic resonances of the air column inside the instrument. The resulting pressure fluctuations $\tilde{p}(t)$ in the mouthpiece amplify the reed oscillations, thus contributing to an increased modulation of the incoming air flow. Therefore, the instability is amplified through a feedback loop. The self-sustained oscillations are expected to occur at a frequency close to a resonance frequency of the air column. However, the reed is a cane plate and has its own mechanical resonances :  undesired squeak sounds may be produced if the instability occurs close to one of these mechanical resonance frequencies (usually the one corresponding to the first flexural mode for the reed). Clarinet players learn to avoid these squeak sounds by properly adjusting two main control parameters : the blowing pressure  inside the mouth, and the force applied by the lower lip on the reed (which alters the position of the reed at rest, thus the height of the reed channel).

The model has been detailed and discussed in~\cite{ica07} and only the resulting fundamental equations are given in this section. For the sake of simplicity, the dynamic behavior of the reed has been omitted. 
Considering as a reference the pressure $p_M=KH$ required to close the reed channel in the non-oscillating case (where $K$ is the reed surface stiffness and $H$ is the height of the reed channel at rest), we introduce the following dimensionless quantities for pressure in the mouth and in the mouthpiece, and for volume flow.
\begin{equation}
p(t)=\frac{\tilde{p}(t)}{p_M}, \quad \gamma=\frac{p_m}{p_M},\quad u(t)=\frac{Z_c \tilde{u}(t)}{p_M},
\end{equation}
where $Z_c=\frac{\rho c}{S}$ is the characteristic impedance for plane waves inside the resonator of cross section $S$, $\rho$ is the air density and $c$ is the sound velocity.
Likewise it is convenient to define a dimensionless reed position :
\begin{equation}
x(t)=y(t)/H+\gamma
\end{equation}
 and $\zeta=Z_cWH\sqrt{2(\rho p_M)^{-1}}$ where $W$ is the width of the reed. The two dimensionless parameters controlled by the player for a given fingering are $\gamma$ and $\zeta$, the blowing pressure and an embouchure parameter respectively.

\paragraph{The air flow}
Mainly based on \textcolor{black}{Hirschberg's considerations}~\cite{Hirschberg:99},  an explicit expression for the air flow $u$ is given below:
\begin{align}
&u=\zeta(1-\gamma+p)\sqrt{|\gamma-p|} sgn(\gamma-p)  & if \quad \gamma-p \leq 1 \label{e:ua}\\
&u=0 & if \quad \gamma-p \geq 1 \label{e:ub}
\end{align}
The first equation corresponds to the case of an open reed channel. In that case, the incoming air flow depends only on the pressure drop $\gamma - p(t)$ between the mouth and the mouthpiece. When the tip of the reed gets in contact with the lay, it completely closes the reed channel, therefore canceling the air flow. This is expressed by the second equation.
In this case, it is noteworthy that the effect of the reed dynamics on the air flow is ignored.
This approximation is valid at low frequency, i.e. when the frequency of the note played is much lower than the first resonance frequency of the reed.

\paragraph{\textcolor{black}{Continuous time equations.}}

We consider a cylindrical bore (with length $L$) for the clarinet. The model is the wave equation inside the tube, with Neumann and Dirichlet boundary conditions at the input and output of the bore respectively, and a source term that accounts for the airflow blown inside the instrument.

Modal coordinates $p_n(t)$ of the mouthpiece pressure  are calculated through the projection on each mode in the time domain. The expansion onto the modes is truncated to the $m$ first modes:
\begin{align}\label{e:ac}
&\frac{d^2 p_1(t)}{d t^2}+\frac{\omega_1}{Q_1}\frac{d p_1(t)}{dt}+\omega_1^2 p_1(t)=\textcolor{black}{\frac{2c}{L}\frac{d u(t)}{dt} }\\ \notag
\vdots\\ \notag
&\frac{d^2 p_m(t)}{d t^2}+\frac{\omega_m}{Q_m}\frac{d p_m(t)}{dt}+\omega_m^2 p_m(t)=\frac{2c}{L}\frac{d u(t)}{dt} \notag \\ \notag
\end{align}
\textcolor{black}{where $\omega_n=(2n-1)\frac{2 \pi c}{4L}$} are the resonance angular frequencies of the air column, and $Q_n$ are the quality factors of the resonances. In equation~\eqref{e:ac}, $u$ is calculated using equations~\eqref{e:ua} and~\eqref{e:ub} with the pressure $p$ calculated as the sum of its modal components: $p(t)=\displaystyle{\sum_{n=1}^m p_n(t)}$. 

Equations (\ref{e:ac}) corresponds to an input impedance for the clarinet defined as:
\begin{equation}
\textcolor{black}{
\label{impe_decom}
Z(\omega)=j\omega \sum_n\frac{2c}{L}\frac{1}{\omega_n^2-\omega^2+j\omega\omega_n/Q_n} 
}
\end{equation}

The quality factors can be derived analytically \cite{chaigne2008acoustique}. Assuming that the radius is small in comparison to the wavelength and the phase velocity is equal to the speed of sound, a good approximation of the quality factors is:
\textcolor{black}{
\begin{equation}
\label{equ:qi}
\frac{1}{Q_n}\approx\frac{2\alpha_1}{R}\sqrt{\frac{\mu L}{\rho \gamma_n c}}+\frac{1}{2}\frac{(k_nR)^2}{\gamma_n}
\end{equation}
where $\gamma_n=\frac{(2n-1)\pi}{2}$, $k_n=\frac{\gamma_n}{L}$, $\mu$ is the viscosity, $\rho$ is the density. $\alpha_1$ is related to the absorption coefficient and is equal to 1.044 in this work (see \cite{chaigne2008acoustique} for details). Unless otherwise stated, the results presented in the remaining of this paper were obtained with a resonator length equal to $L=1 \text{m}$, an interior radius equal to $R = 7 \text{mm}$, a viscosity $\mu=1.708 \times10^{-5}$ kg.m$^{-1}$s$^{-1}$, and a density $\rho=1.29$ kg.m$^{-3}$.}

\subsection{Numerical Implementation}

The system of equations~\eqref{e:ac} is solved using \textsc{ode} solvers. For this purpose, the second order \textsc{ode}s are reformulated into a system of $2m$ first order \textsc{ode}s. As multiple time scales are involved in the problem (duration of blowing pressure transient, bore resonance frequencies, reed resonance frequency and reed-lay contact duration), an \textsc{ode} solver designed for stiff problems, namely \texttt{ode15s} from the Matlab \textsc{ode} Suite was used. Note that in order to obtain consistent and robust pressure time series, the relative and absolute convergence tolerances were tightened to $10^{-5}$ and $10^{-7}$ respectively.

\section{\textcolor{black}{KEY FEATURES OF THE APPROACH IN PRACTICE}}

\textcolor{black}{
In this section, the key features and potential of the approach are presented in the context of musical acoustics. These features are demonstrated through several test cases based on the simplified single reed instruments model described in the previous section.}



\subsection{\textcolor{black}{Efficiency of the adaptive sampling}}

\textcolor{black}{In order to highlight the efficiency of the adaptive sampling,} the first set of results is based on the separation between regimes producing sounds and the soundless ones. We first use a model of clarinet with one mode (Equation~\eqref{e:ac} with $m=1$).

\begin{figure}
\begin{minipage}{0.47\linewidth}
\centering
\includegraphics[width=1\textwidth]{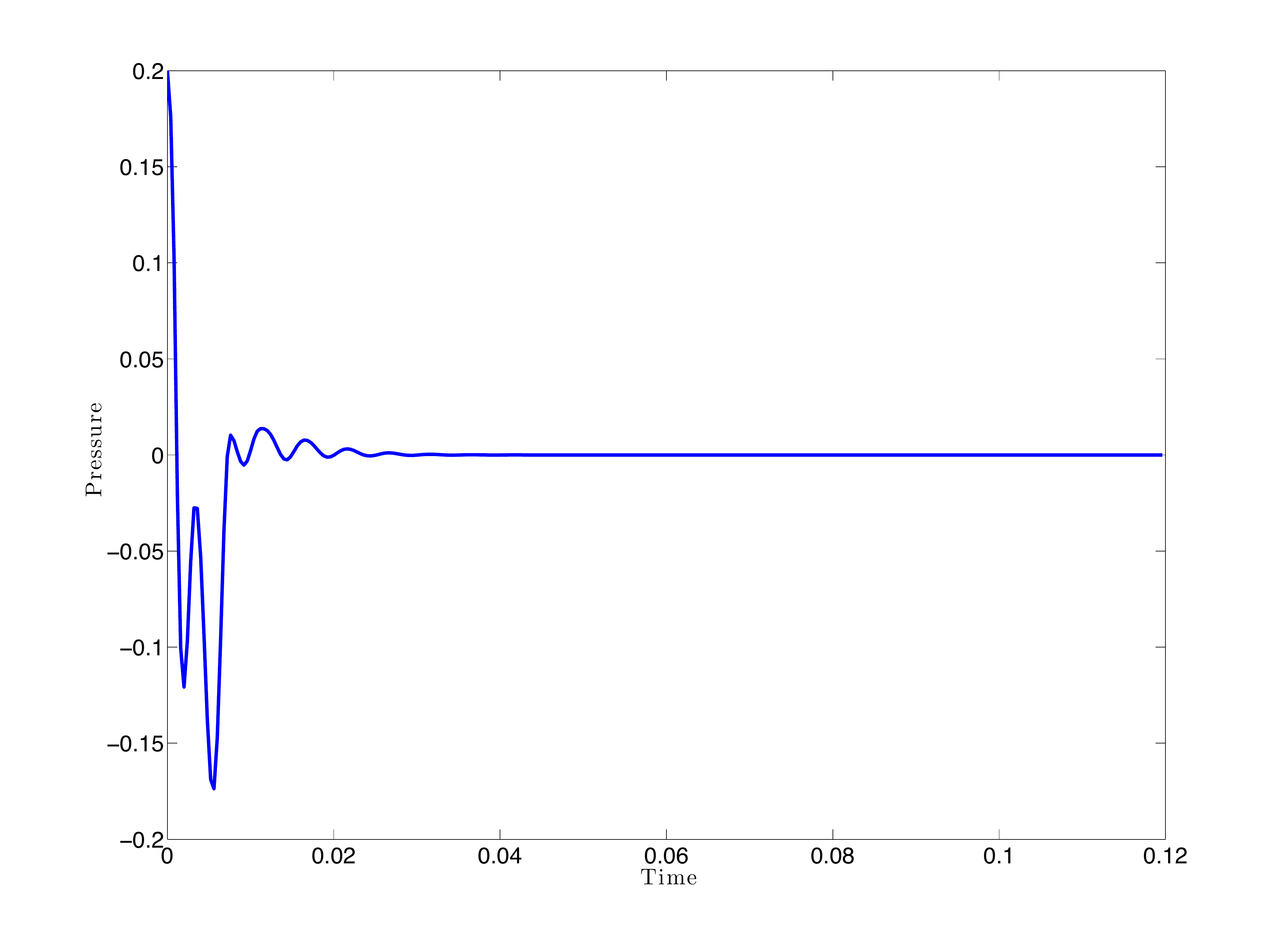}
\caption{Soundless configuration. Static regime reached after a transient.\label{fig_nosound}
}
\end{minipage}
\begin{minipage}{0.02\linewidth}
$\phantom{a}$
\end{minipage}
\begin{minipage}{0.47\linewidth}
\centering
\includegraphics[width=1\textwidth]{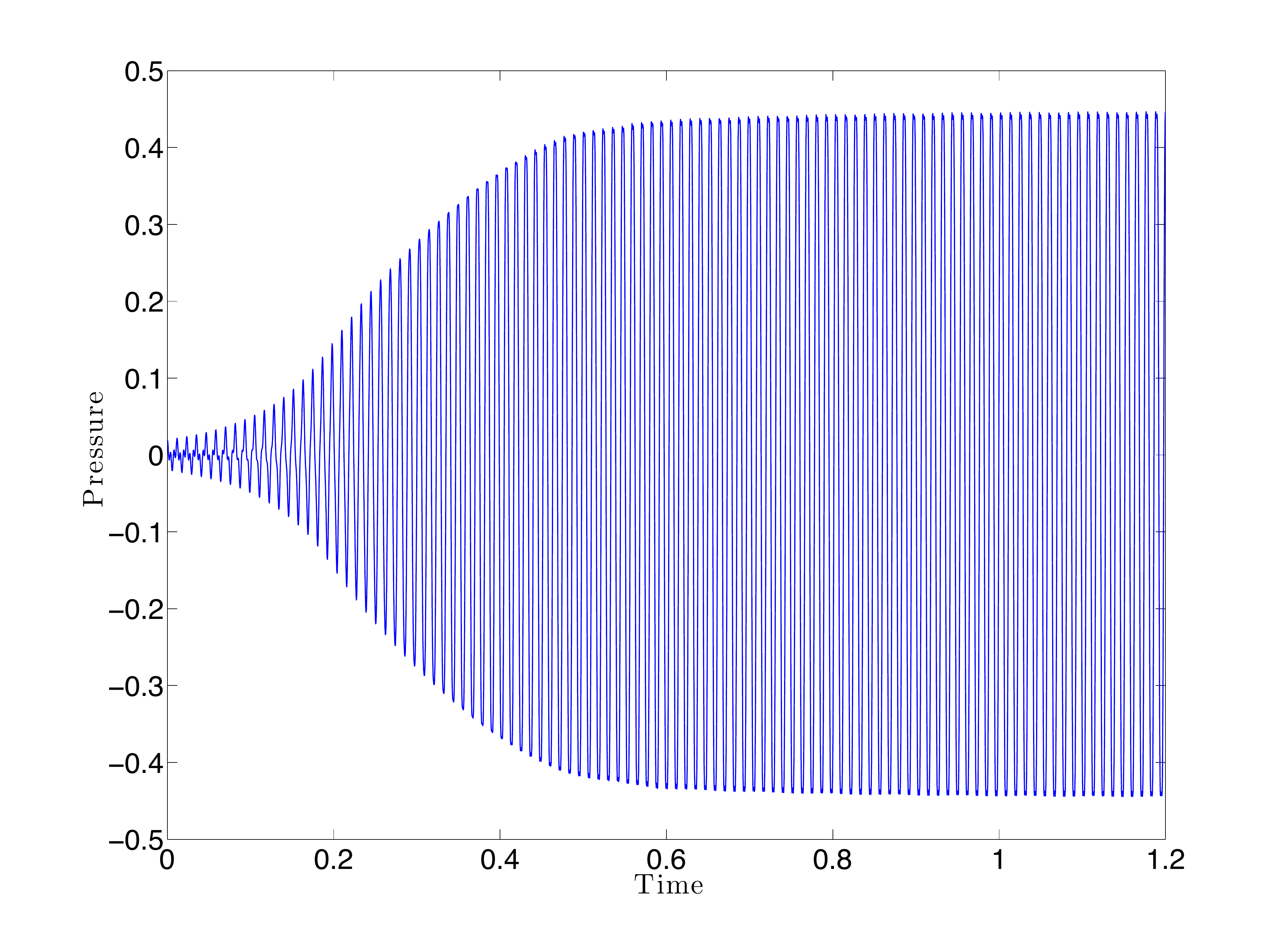}
\caption{Configuration producing a sound through self-sustained oscillations.\label{fig_sound}}
\end{minipage}
\end{figure}

Soundless configurations are obtained if the initial oscillations die down and reach a static regime, as shown in Figure~\ref{fig_nosound}. On the other hand a sound is produced if self-sustained auto-oscillations are reached as depicted in Figure~\ref{fig_sound}.
In order to differentiate between the two states, a simple tentative criterion is used: the oscillation amplitude is averaged over the last third $N_{2/3}$ of the time series. For a large enough time, the average value is compared to an arbitrary value $\epsilon_1$.  Formally, the criterion reads:

\begin{align}
\frac{1}{N_{2/3}}\sum_{N_{2/3}}p(t_i) > \epsilon_1 & \quad \text{Oscillations (i.e., sound)} \\ \notag
\frac{1}{N_{2/3}}\sum_{N_{2/3}}p(t_i) \leq \epsilon_1 & \quad \text{Static regime (i.e., no sound)} . \\ \notag
\end{align}

\begin{figure}
\begin{minipage}{0.50\linewidth}
\centering
\includegraphics[width=0.99\textwidth]{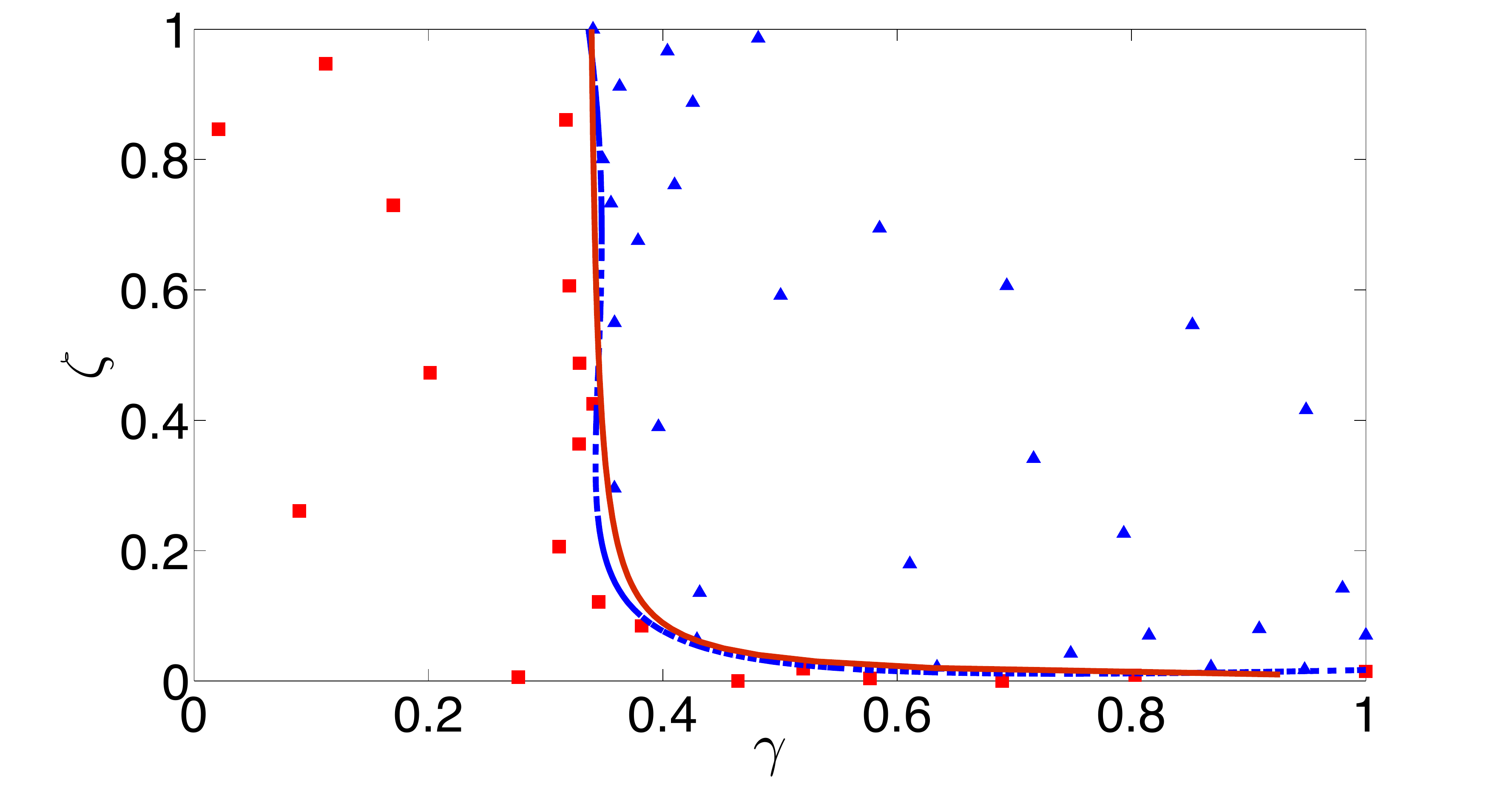}
\centering \tiny{a) 20 DOE + 30 adaptive samples}
\end{minipage}
\begin{minipage}{0.50\linewidth}
\includegraphics[width=0.99\textwidth]{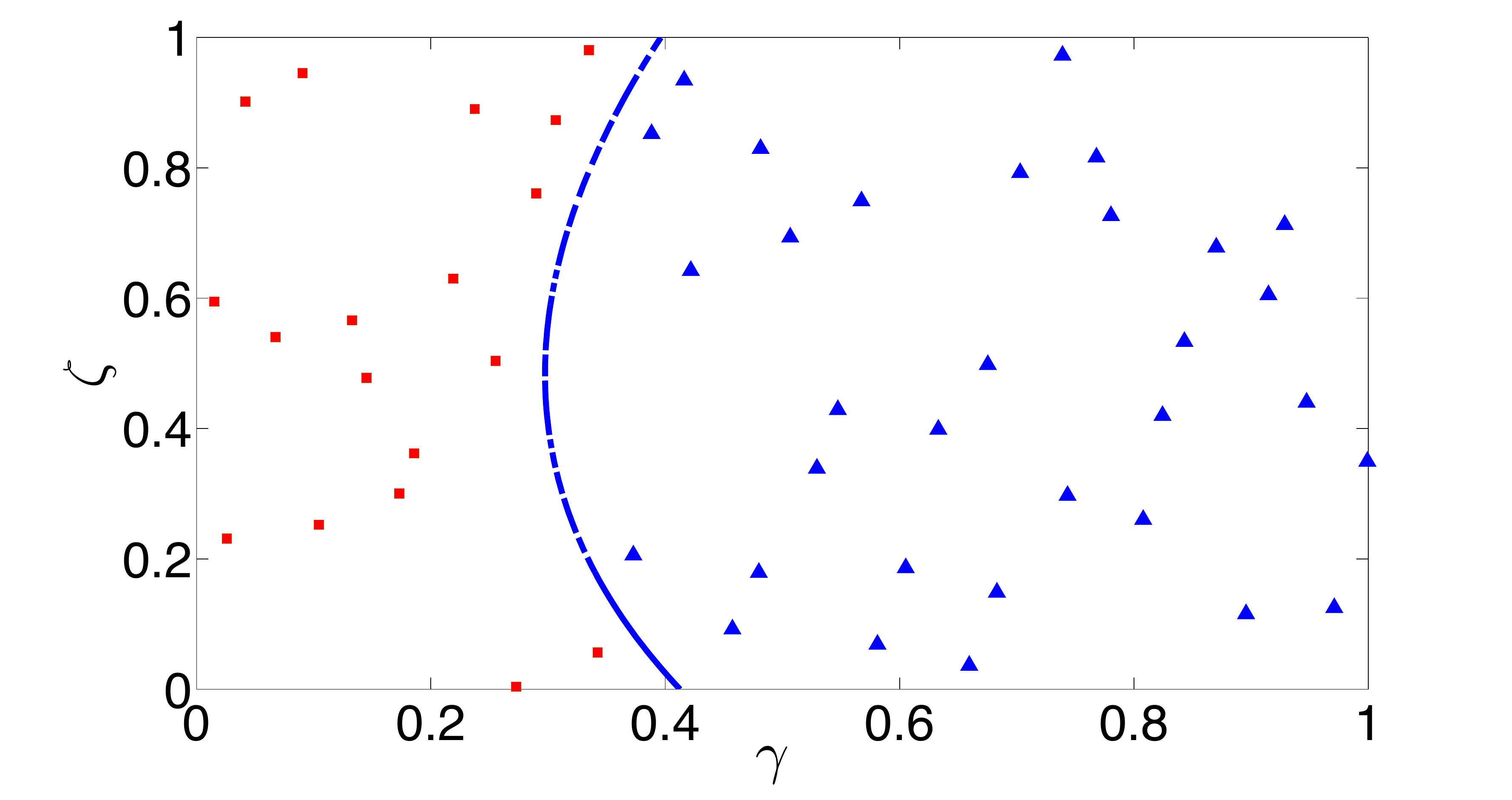}
\centering \tiny{b) 50 DOE samples}
\end{minipage}
\caption{Boundary obtained using SVM with adaptive sampling (dashed blue curve) with a total of 50 samples (a). The SVM boundary separates sound producing regimes (blue triangles) and soundless ones (red squares). The solid red line is the theoretical boundary \cite{chaigne2008acoustique}. For comparison, boundary created with 50 samples without adaptive sampling (b).}
\label{fig_res_sound2d}
\end{figure}

The first boundary is constructed in the ($\zeta,\gamma$) space (adimensional reed channel opening and mouth pressure). For a given instrument design, these two parameters can be considered as the most influential control parameters. Figure \ref{fig_res_sound2d} depicts the corresponding boundary constructed using 20 DOE samples and 30 adaptive samples.

\begin{figure}
\centering
\includegraphics[width=0.65\textwidth]{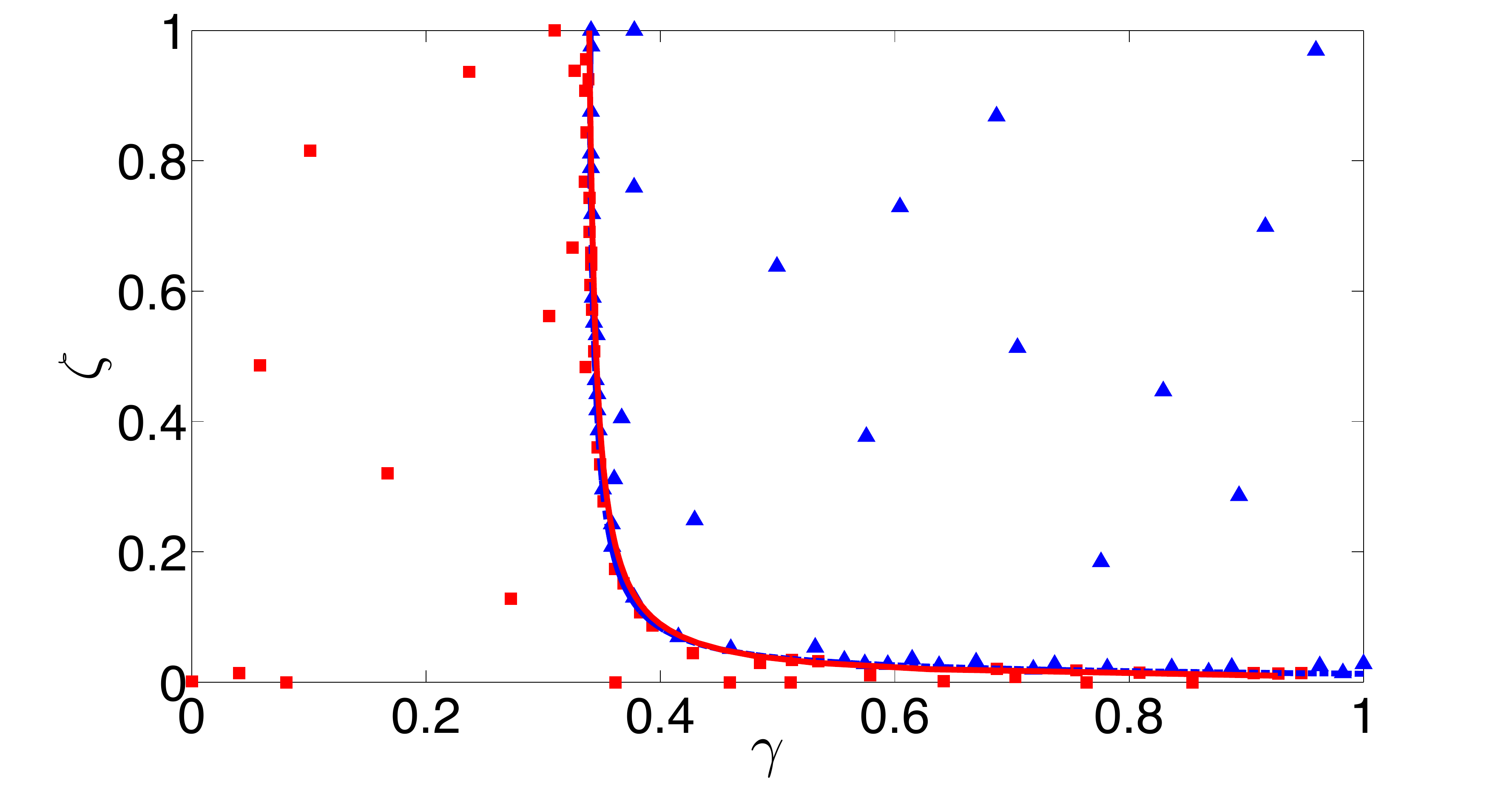}
\caption{Approximated and theoretical boundaries. 20 initial DOE and 80 adaptive samples. The SVM and theoretical boundaries are almost identical.}
\label{fig_res_sound2d_2}
\end{figure}

The theoretical threshold for the soundless configurations can be derived analytically  \cite{chaigne2008acoustique}:
\begin{equation}
\label{eq_threshold_theory}
\gamma_{th}=\frac{1}{3}+\frac{\alpha L}{\zeta}\frac{2}{3\sqrt{3}}
\end{equation}
where $\alpha=\frac{\pi^2}{16QL}$. $Q$ is the quality factor set to 40 for this problem. The corresponding boundary is superimposed to the SVM approximation in Figure~\ref{fig_res_sound2d}. In the case a total of 50 samples were used (20 DOE samples and 30 adaptive), there is a small difference between the theoretical and SVM boundary. This difference disappears in the case when 80 adaptive samples are used as showed in  Figure \ref{fig_res_sound2d_2}.

This result reveals an important and well known conclusion: for most of the reed channel opening (represented by $\zeta$), a sound is obtained if the pressure in the mouth (represented by $\gamma$) is approximately larger than one third. For small values of $\zeta$, this is no longer valid with a threshold asymptotically increasing.

The example is also used to show the importance of the adaptive sampling scheme. For this purpose, the same boundary was created using a static design of experiments with 50 samples  (i.e., no adaptivity), thus leading to a rather inaccurate boundary (Figure \ref{fig_res_sound2d} b). The advantage of the adaptive sampling scheme was shown to become more pronounced as the dimension increases \cite{basudharimproved}.

\begin{figure}[h]
\centering
\includegraphics[width=0.75\textwidth]{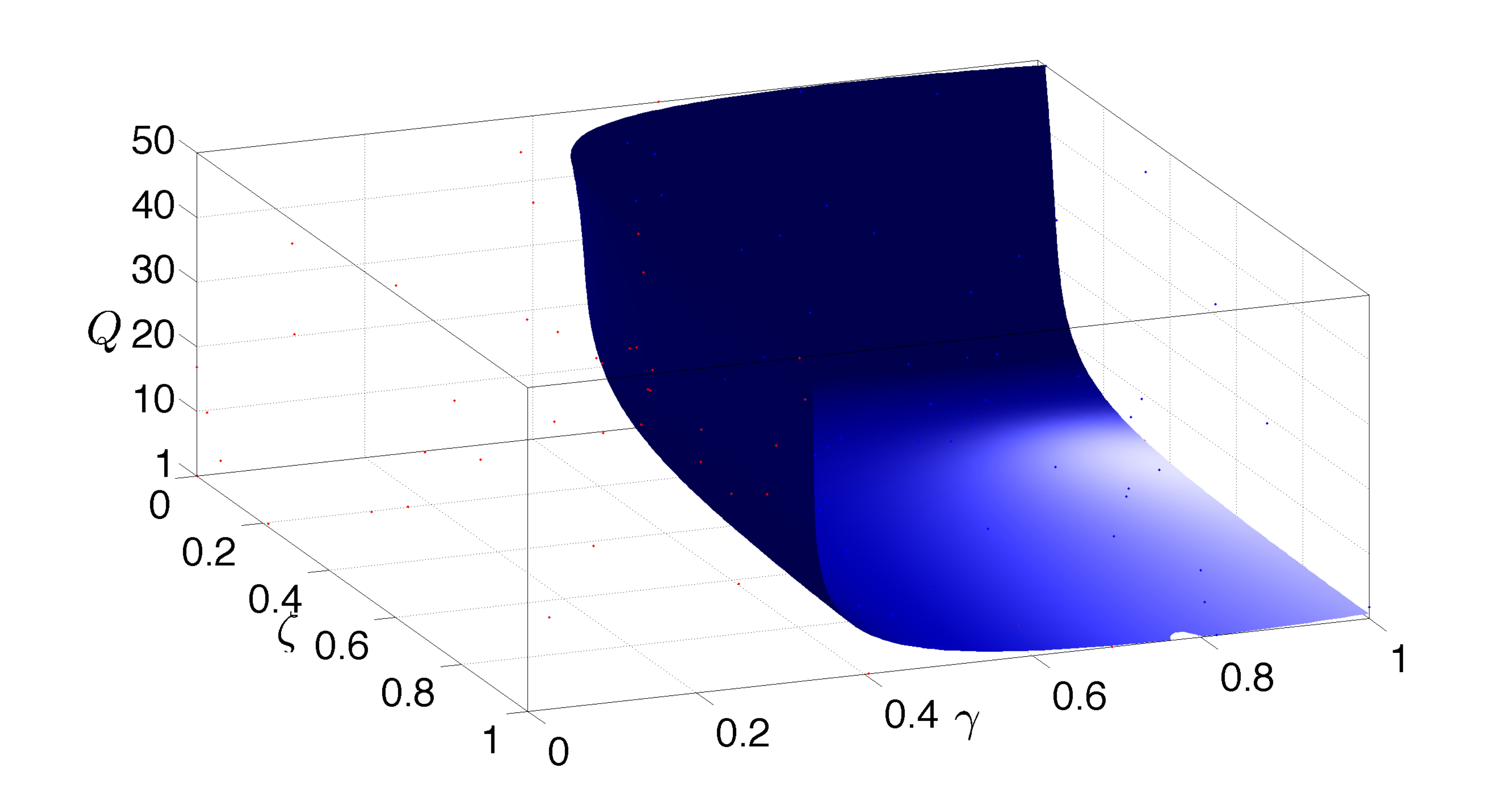}
\caption{Boundary constructed in the ($\zeta,\gamma, Q$) space. The SVM boundary separates sound producing regimes (blue dots) and soundless ones (red dots). \label{fig_sound_3D}
}
\end{figure}

The second boundary is constructed in the ($\zeta,\gamma, Q$) space  (Figure \ref{fig_sound_3D}), which, in this case, is assumed to be freely imposed. The boundary is based on 50 DOE and 100 adaptive samples. As expected, it shows that the lower the quality factor is, the higher the oscillation threshold. This is due to the fact that more losses in the resonator should be compensated by an additional amount of energy (i.e. a larger blowing pressure).

Through the two previous simple problems, it appears that the approach proposed in this paper could be used as a new way of studying the oscillation threshold in musical instruments. Note that other approaches in the literature use analytic developments around the static solution~\cite{wilson74,silva2008b,cullen2000a}, numerical continuation of Hopf points~\cite{karkar2012} or regular sampling of the design space~\cite{inacio2007}. The proposed method generalizes these attempts by removing the restriction on the types of models (black-box models can be used) and by providing a sampling strategy more effective than a simple grid approach. 

\subsection{\textcolor{black}{Ability to converge towards complex boundaries}}

\textcolor{black}{To demonstrate the ability of the method to generate complex boundaries,} this set of results provide examples related to the playing frequency. 
The frequency criterion is based on the difference between the actual frequency of the signal $f_{act}$ and a reference frequency $f_{ref}$. It is typically expressed in cents. The number of cents $N_{cents}$ is defined as:
\begin{equation}
N_{cents}=1200log_2 \left( \frac{f_{act}}{f_{ref}} \right) 
\end{equation}

The criterion chosen  compares $N_{cents}$ to a user defined threshold $\epsilon_2$. The reference frequency can be the frequency of a note (e.g., 440Hz). In our work we have used the natural frequencies of the resonator as reference or notes from the equal-tempered scale. The actual frequency of the signal was obtained using the freely available code Yin \cite{yin}.

\begin{figure}
\centering
\includegraphics[width=0.8\textwidth]{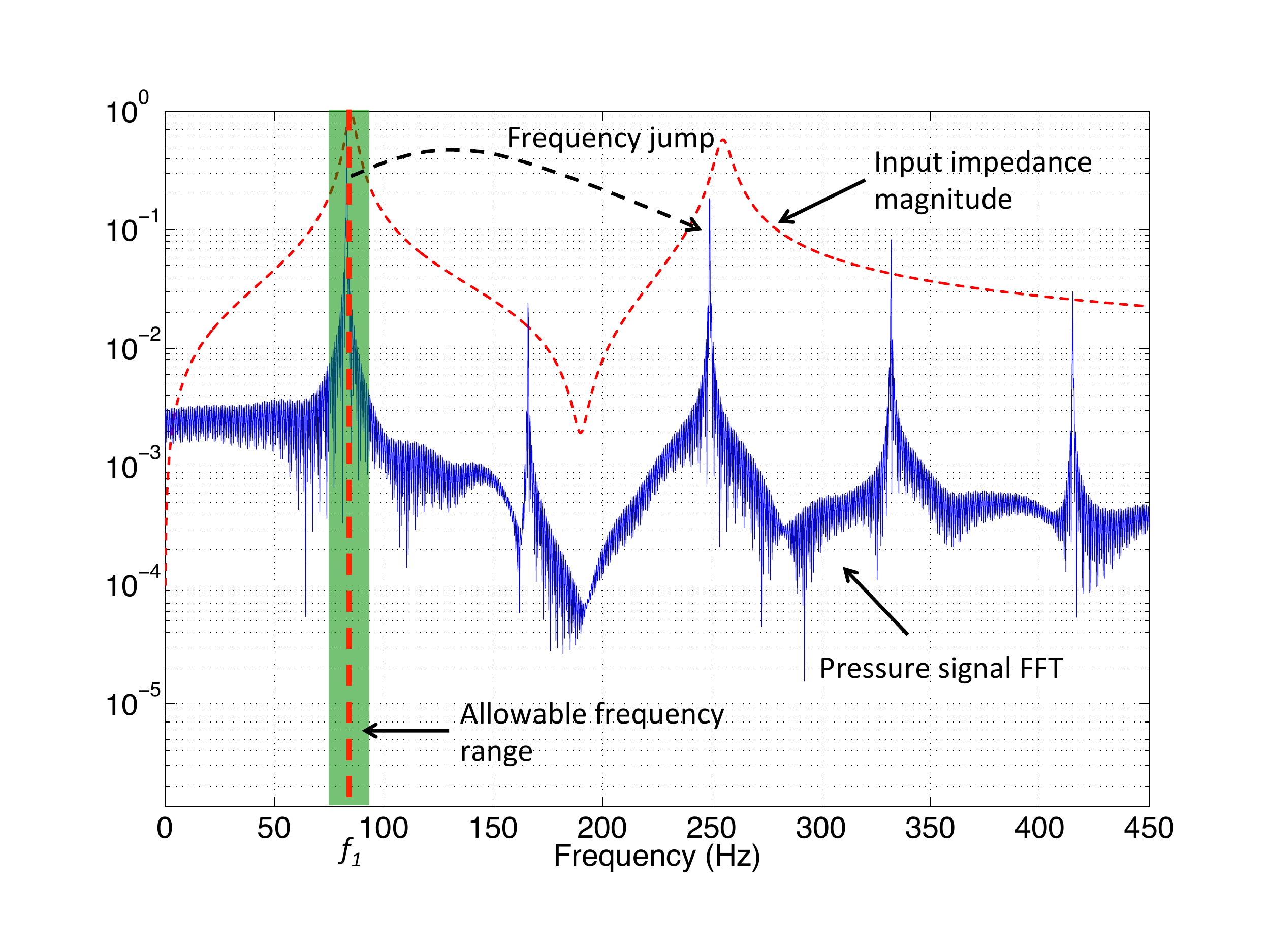}
\caption{Depiction of the frequency criterion. The light green strip represents acceptable playing frequencies with a reference frequency chosen as $f_1$. Possible jump from one peak of the frequency response function to another, thus leading to a discontinuity.\label{fig_twopeaks}
}
\end{figure}

It is noteworthy that the frequency of the produced sound can, for instance,  jump from one neighborhood of the resonators frequencies to another. That is, the signal frequencies can be discontinuous, thus requiring specific approaches such as the classification technique proposed for the construction of the maps. Figure (\ref{fig_twopeaks}) depicts the frequency criterion along with possibility of a ``jump".

The criterion is applied to the clarinet model with two modes (Equation~\eqref{e:ac} with $m=2$). The quality factors where approximated using analytical expressions available in Equation~\eqref{equ:qi}. The criterion aims at separating the configurations within a given number of cents from a specific frequencies (e.g., the resonator's natural frequencies or a given pitch) and the other configurations,  including the static regime configurations (i.e., soundless ones).  As pointed previously, it is worth insisting on the fact that the use of a classification technique is justified by the fact that the frequencies are discontinuous. 

In this experiment, the criterion is used to identify the playing frequencies that are within 5 cents of the first natural frequency of the resonator.  The natural angular frequencies of the resonator are:
\begin{equation}
\omega_k=\frac{ q \pi c}{2L} \quad \quad q=1,3,5,..
\end{equation}

\begin{figure}
\centering
\includegraphics[width=0.75\textwidth]{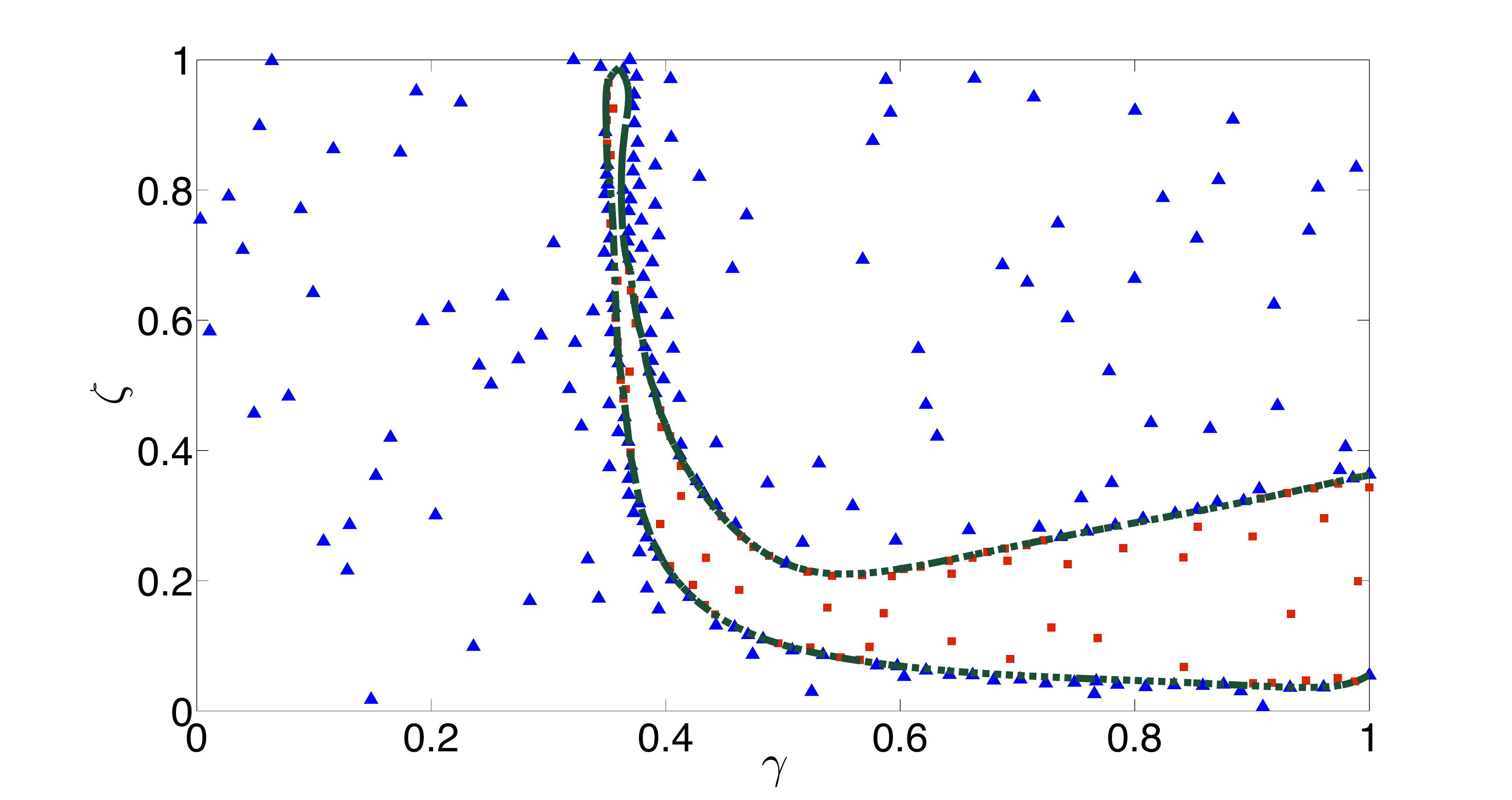}
\caption{SVM boundary identifying configurations within 5 cents of the first resonator natural frequency (red squares). The region of interest is surrounded by static solutions and the solutions with a difference larger than 5 cents (blue dots).\label{fig_freq}
}
\end{figure}

For a resonator of length $L=1\text{m}$, this leads to a first and second natural frequency of  85.75Hz and 257.25Hz respectively. Figure~\ref{fig_freq} depicts the corresponding map constructed from a design of experiments with 100 samples and 180 adaptive samples generated with the scheme described in Section~\ref{sec:EDSD}. It is noteworthy that despite the apparent simplicity of the map, its accurate construction with a reasonable number of samples is highly dependent on the initial design of experiments. Specifically, in the narrow region of the map corresponding to large values of the opening, if no sample is initially present in this area, the adaptive sampling scheme would have difficulty (i.e., would exhibit a poor convergence rate) to construct the boundary.

Figure~\ref{fig_freq} shows that the reed opening $\zeta$ partly dictates the ease of play of a given note.
If the opening is too large (little support of the reed by the lip, i.e. high values of $\zeta$), the blowing pressure $\gamma$ must be very precisely controlled to play in tune.
On the contrary, when the lip reduces the opening of the reed channel, playing in tune becomes easier for the musician because a lower accuracy is required on the control of the blowing pressure.
\textcolor{black}{To the authors' knowledge, this has rarely been addressed in the literature with the exception of a very recent work \cite{CoyleSMAC2013}}.

\begin{figure}[h]
\centering
\includegraphics[width=0.75\textwidth]{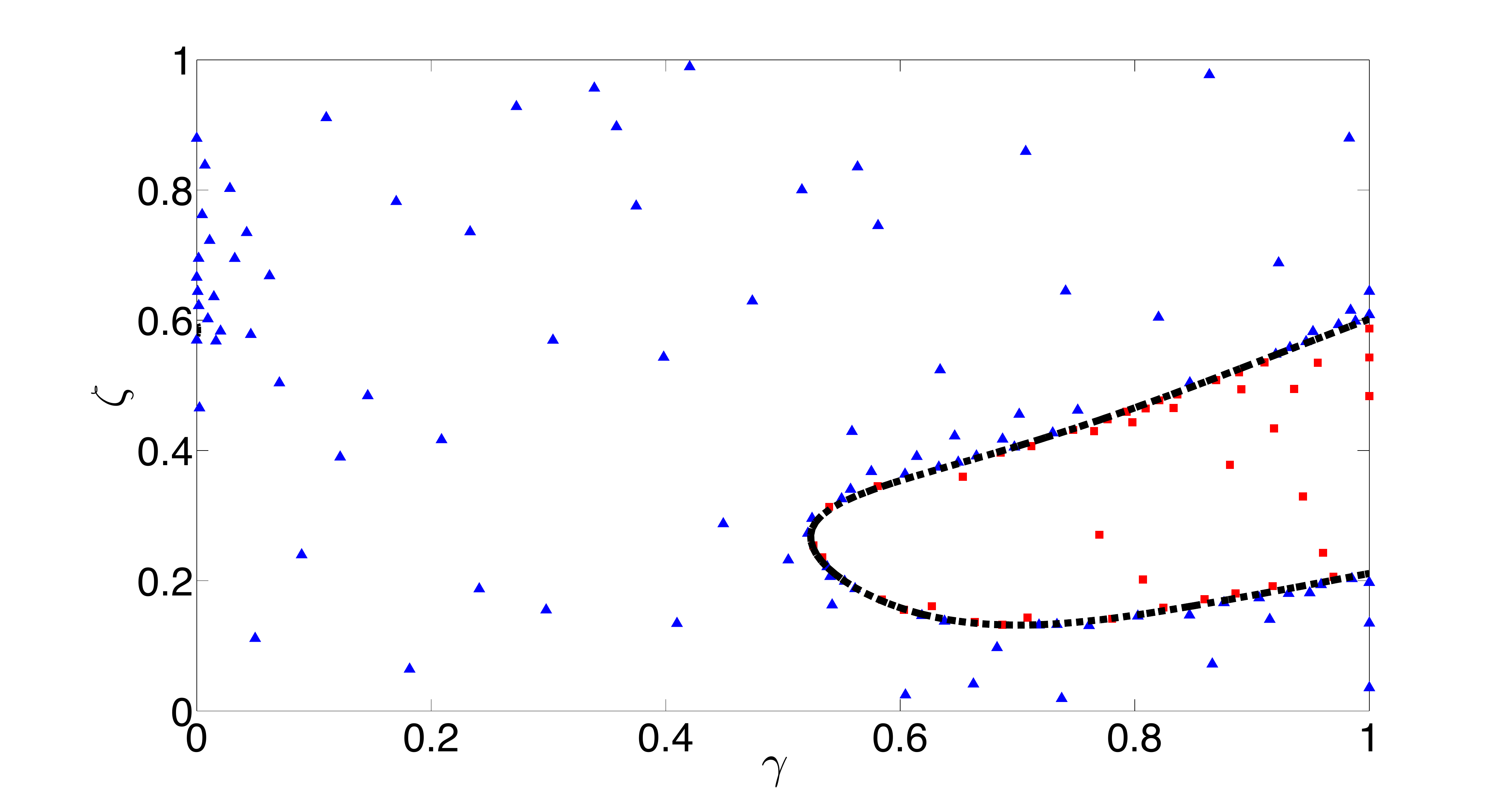}
\caption{SVM boundary identifying configurations within 5 cents of the second register (red squares).\label{fig:secreg}
}
\end{figure}

The same experiment is repeated for the second register. In order to increase the number of cases reaching the second register (i.e., the second natural frequency of the resonator), the initial condition of the second component of the adimensional pressure in the modal basis was set to 0.9 while the first component was set to 0.09. Figure \ref{fig:secreg} depicts the region of the $(\gamma, \zeta)$ space where the playing frequency is within 5 cents of the second register.  The map was constructed using 50 initial DOE samples and 100 adaptive samples.

In order to demonstrate the flexibility of the approach and study the influence of initial conditions, a variable was introduced whose purpose is to weigh differently the initial condition of the first and second components of the pressure in the modal basis. The weight $\beta$ is chosen so that $p_1(t=0)=\beta$ and $p_2(t=0)=1-\beta$. This represents of course an artificial problem whose only purpose is to demonstrate the various types of parameters that can be involved in the construction of the maps. On a side note, the ability to control the registers to be reached can be useful for sound synthesis. 

\begin{figure}
\centering
\includegraphics[width=0.85\textwidth]{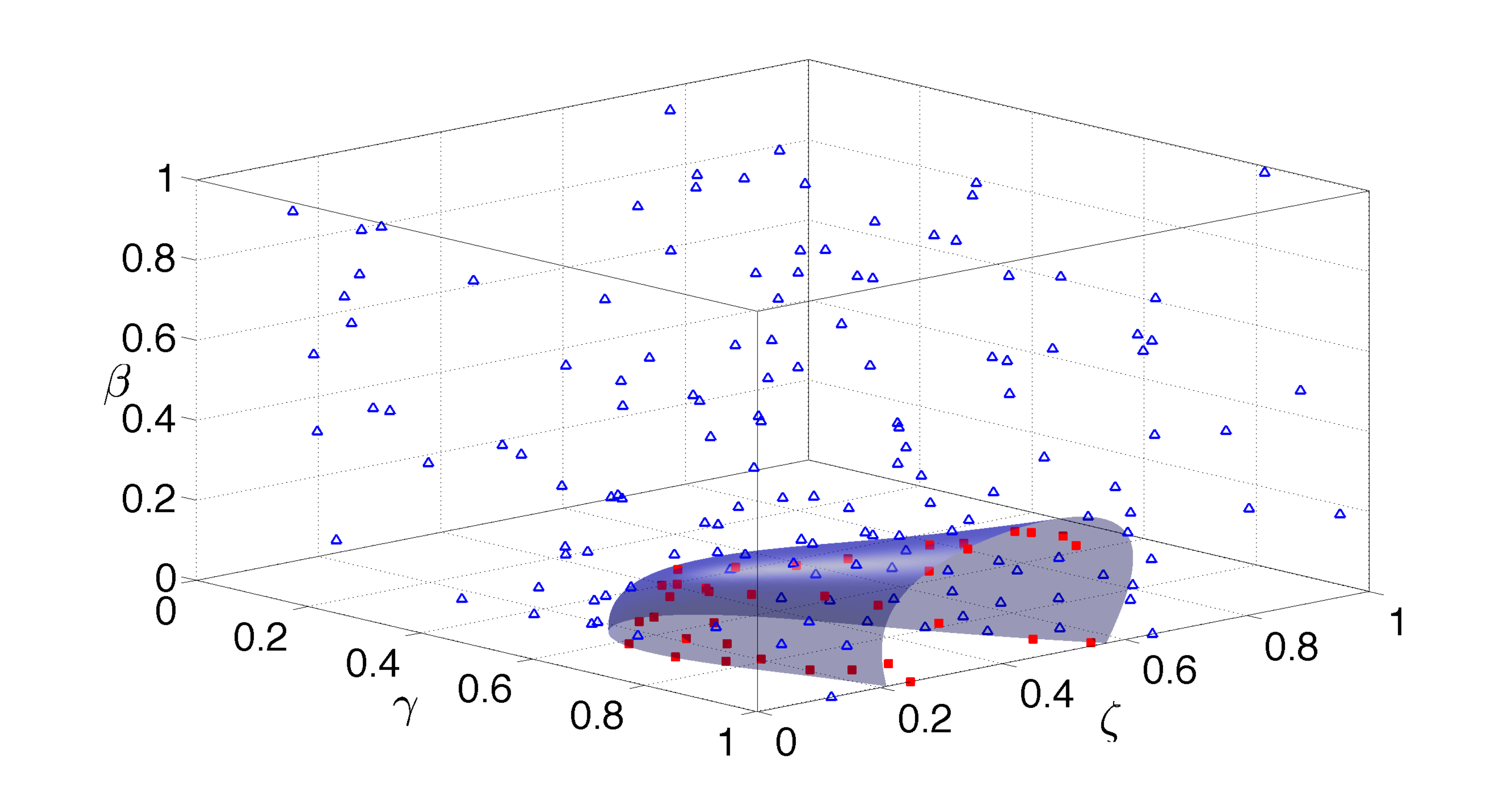}
\caption{Region of the space within 5 cents of the second register (red squares). Inclusion of initial conditions through the parameter $\beta$. \label{fig:inisecondreg}
}
\end{figure}

In a three-dimensional space $(\beta,\gamma,\zeta)$, we wish to identify the region of the space leading to the second register within 5 cents as in the previous problem. Figure~\ref{fig:inisecondreg} depicts the results obtained with 100 initial DOE samples and 80 adaptive samples.  As expected, when the initial condition on $p_2$ increases (i.e., $\beta$ decreases), the region of the space within 5 cents of the second register also increases. 

Figure~\ref{fig:inisecondreg} highlights the strong impact of initial conditions on the behavior of the dynamical system.
If $\beta$ is between 0.4 and 1, it is impossible to play in tune on the second register regardless of the parameter values $\gamma$ and $\zeta$.

\subsection{\textcolor{black}{Probability estimates}}

As explained in Section~\ref{sec:prob},  SVM can be used for the calculation of probabilities of belonging to either the ``positive" or ``negative" region of the space. Given probability density functions for the parameters, the probability of belonging to a region of interest can be efficiently computed using Monte-Carlo simulations with Equation \ref{eq:MCprob}.

\begin{figure}
\centering
\includegraphics[width=0.70\textwidth]{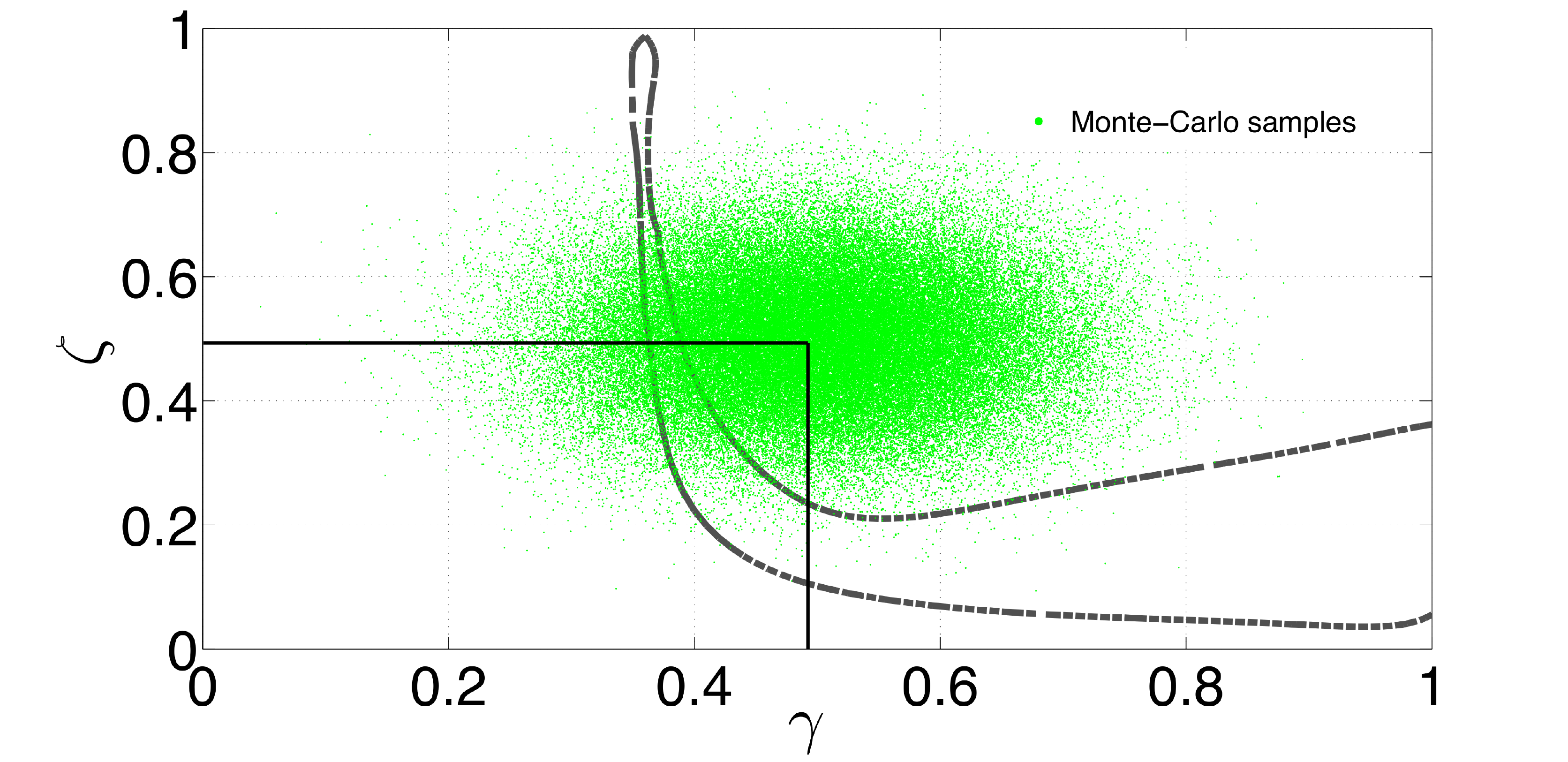}
\caption{Example of probability estimates of belonging to the region where the playing frequency is within 5 cents of the first natural frequency of the resonator.\label{fig:exprob}
}
\end{figure}

\textcolor{black}{A probabilistic approach is particularly interesting in the context of analyzing or designing musical instruments since control parameters and design parameters follow statistical distributions and are subjected to uncertainties. For instance, the blowing pressure might be assumed to follow a specific statistical distribution (e.g., normal) or the manufacturing tolerances might introduce uncertainties that with influence the acoustic behavior of the instrument}. As an example, consider the map depicted in Figure \ref{fig:exprob}  describing the region where the playing frequency is within 5 cents of the first resonator frequency.

\textcolor{black}{We wish to evaluate the probability of satisfying the intonation criterion knowing that the parameter $\gamma$ and $\zeta$ follow normal distributions with identical means and standard deviations: $\gamma \sim N(0.5, 0.05)$ and $\zeta \sim N(0.5, 0.05)$. Figure \ref{fig:exprob} depicts the corresponding joint distribution over the map with $10^5$ Monte-Carlo samples. The resulting probability of belonging to the region with an ``acceptable frequency" is 6.3\%.} To the authors knowledge, it is the first time that uncertainty is introduced in the investigation of tuning or playability.

\subsection{\textcolor{black}{A simple example of optimization for intonation}}

\begin{figure}[h]
\centering
\includegraphics[width=0.75\textwidth]{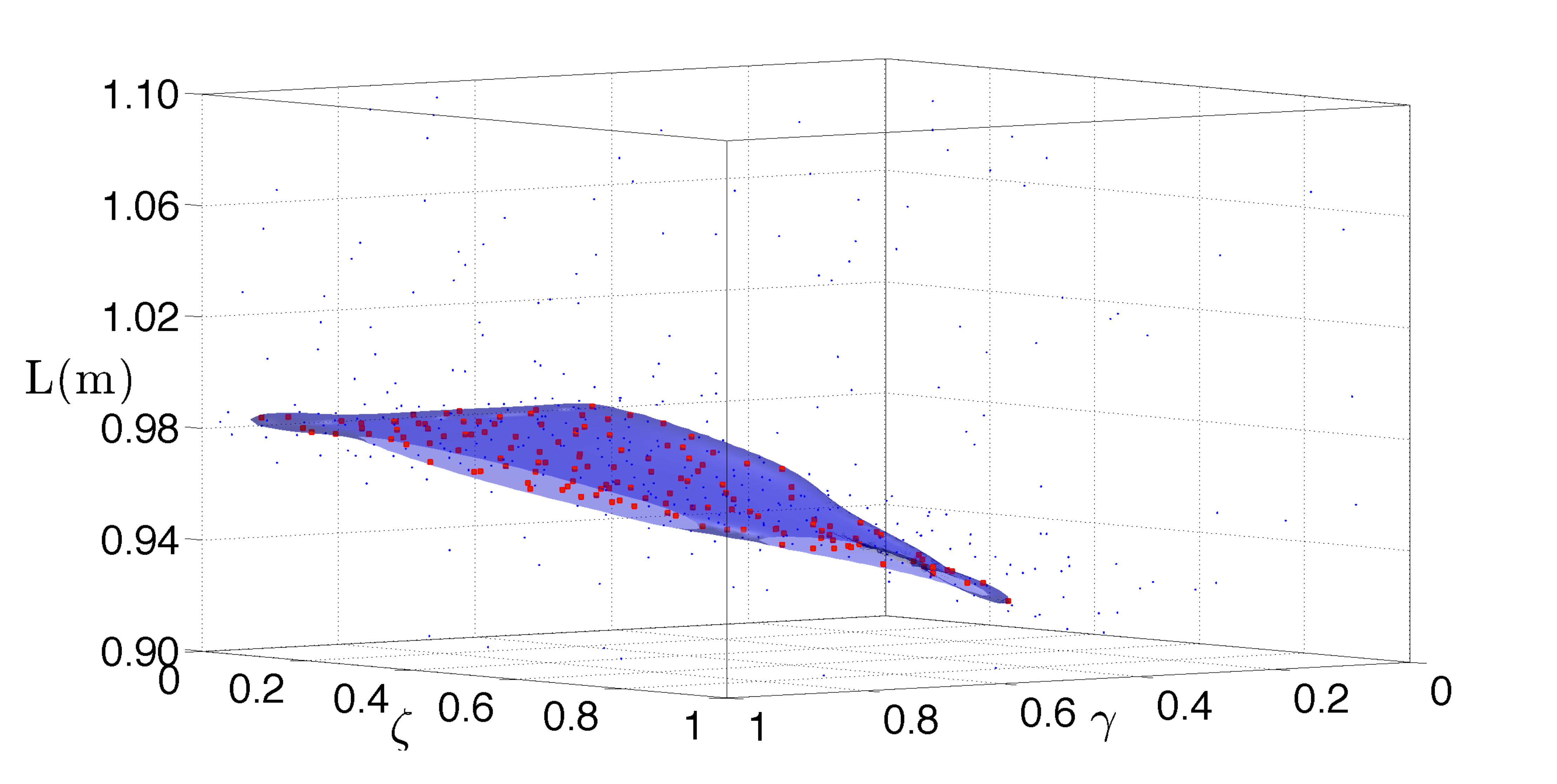}
\caption{Intonation map with respect to the length, $\gamma$ and $\zeta$. The reference frequency corresponds to the note F2 of the well-tempered scale.\label{fig:exdesign}
}
\end{figure}

The ability to generate maps and to assess the probability of belonging to specific regions provides a framework to perform design optimization. As a demonstrative example, we wish to find the optimal length of a resonator so as to maximize the probability of having a playing frequency within 5 cents of the given note from the twelve-tone equal temperament scale.
This ``design for intonation" is of major practical importance in the world of musical instruments. As an example, consider the closest frequency in the equal temperament scale to the first natural frequency of a 1 m long resonator (85.75Hz). The corresponding frequency is found to be 87.30 Hz (note F2). The objective is to find the optimal length $L^*$ of the resonator that will maximize the probability of playing F2 within $+/-$ 5 cents as a function of $\gamma$ and $\zeta$. Formally, this can be written as:

\begin{equation}
L^*=\underset{L}{\operatorname{argmax}} \quad P((\gamma,\zeta) \in \Omega),
\end{equation}
where $\Omega$ is the region satisfying the intonation criterion. The region of interest is depicted in Figure \ref{fig:exdesign}. It was constructed using 100 DOE samples and 400 adaptive samples. Note that the region is rather thin and would not be easily obtainable without adaptive sampling.

\begin{figure}[!h]
\begin{minipage}{0.5\linewidth}
\centering
\includegraphics[width=0.99\textwidth]{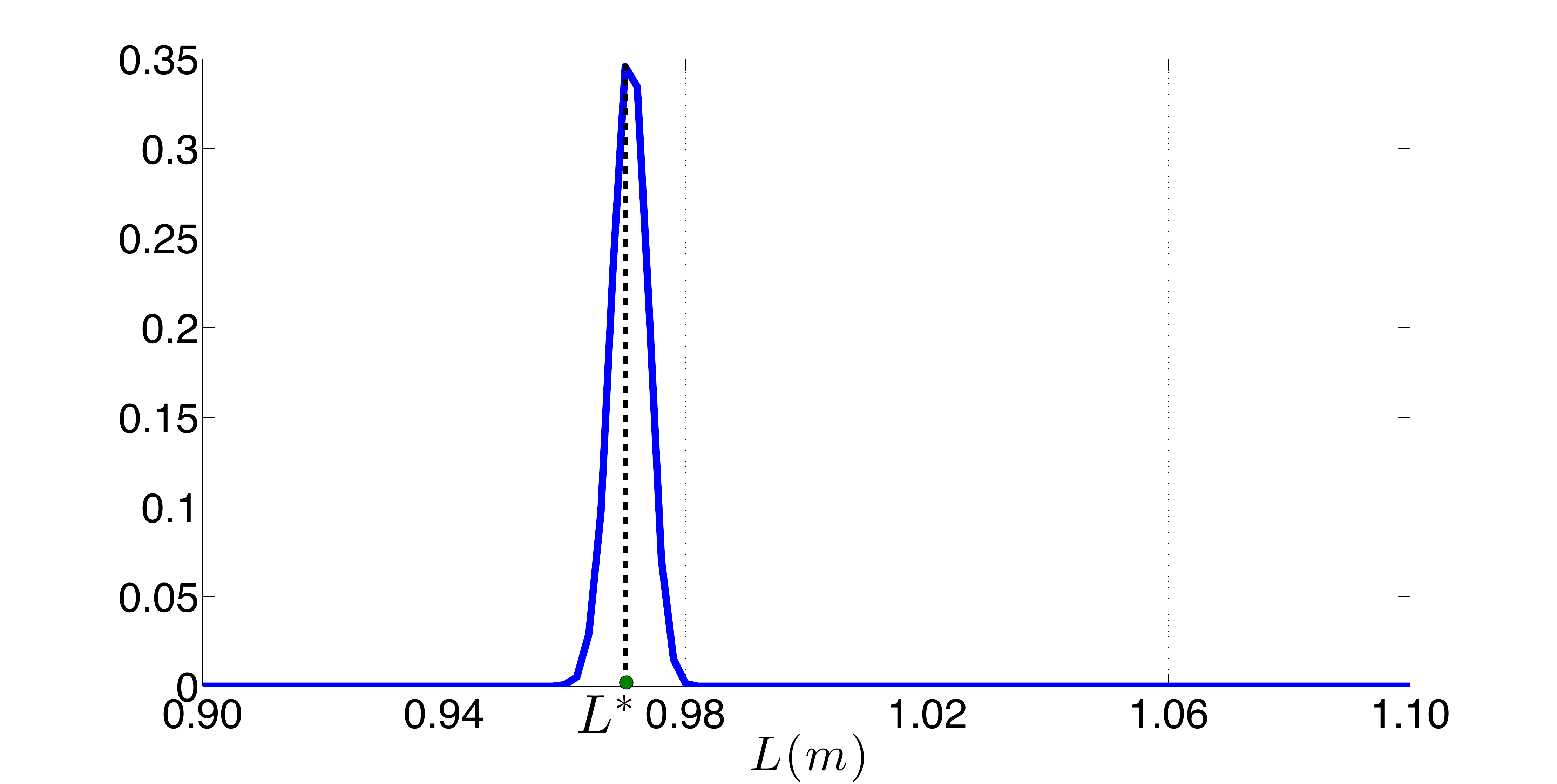}
\centering \tiny{a)} 
\end{minipage}
\begin{minipage}{0.5\linewidth}
\centering
\includegraphics[width=0.99\textwidth]{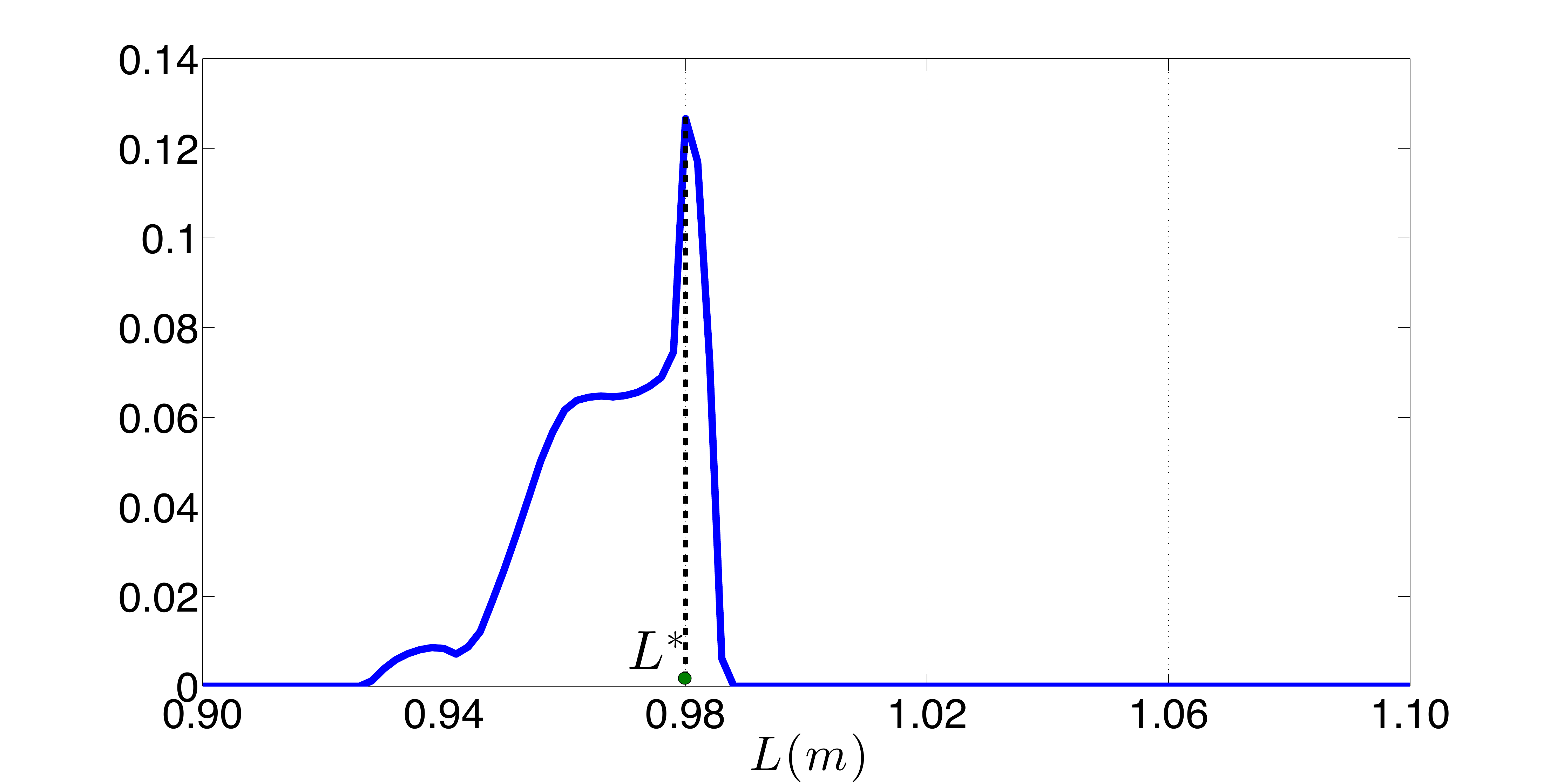}
\centering \tiny{b)} 
\end{minipage}
\caption{Probability of ``good" intonation as a function of the length. Case of normally distributed (a) and uniformly distributed (b) $\gamma$ and $\zeta$. \label{fig:prob}}
\end{figure}

For a given value of the length, the probability can be estimated using  Monte-Carlo simulations as in Section \ref{sec:prob}. In this work, the optimal value of $L$ is searched within the interval [0.9, 1.10] m. Two cases of probability distributions are used for $\gamma$ and $\zeta$: normal $N(0.5,0.05)$ and uniform $U(0,1)$. The corresponding probabilities as a function of the length are depicted in Figure \ref{fig:prob} as well as the corresponding optima. The optimal length is 0.97 m  in the normally distributed case and 0.98 m for uniformly distributed variables.

Because this is a one dimensional optimization problem, the optimal length can be obtained by testing an array of cases. For higher dimensional optimization problems (including for instance side hole positions and sections), dedicated optimization techniques should be used. 

\section{CONCLUSION}
\textcolor{black}{A methodology was proposed to map the acoustic behaviors of wind instruments. 
A map of the acoustic regimes is accurately constructed using a technique referred to as explicit design space decomposition. This approach, based on an SVM classifier, enables one to accurately define \emph{explicit} boundaries of the region of interest in terms of the selected parameters (e.g., blowing pressure and reed opening).}


\textcolor{black}{The results demonstrate the ability of the proposed approach to provide an understanding of the coupling between the parameters that could not be obtained otherwise. This difficulty stems from the highly nonlinear nature of the sound production in wind instruments. The maps produced for the two criteria are representative of the complexity of the phenomenon and the strong coupling between the parameters. In addition, these examples also demonstrate the potential high sensitivity of the acoustic behavior to small perturbations and uncertainties that the proposed technique is able to propagate. Finally, beyond analysis, this paper has demonstrated that the proposed mapping can be used as a tool for design. To the authors knowledge, this represents the first attempt to generate such maps.}

\textcolor{black}{The next steps of the research will include studies in higher dimensional spaces with more accurate models. In addition, the approach will also be tested to generate maps obtained experimentally with an artificial mouth~\cite{Almeida,Lopes} using a reasonable number of measurements.  Finally, the mapping technique will be used to attempt instrument design optimization.}

\section{ACKNOWLEDGMENTS}
This study was performed during a visit of the first author to the CNRS in Marseille, France.  This work was performed within the framework of the Labex MEC (ANR-11-LABX-0092) and has received support from the french National Research Agency (project AMidex (ANR-11-IDEX-0001-02), and project CAGIMA (ANR-11-BS09-0022)). The support of the National Science Foundation for the explicit design space decomposition part (grant CMMI-1029257) is gratefully acknowledged. The authors are also indebted to Dr. Jean Kergomard for priceless technical discussions.

\bibliographystyle{model1-num-names}
\bibliography{BIBJSV}

\end{document}